\documentclass[aps,prl,twocolumn,floatfix,superscriptaddress]{revtex4-1}
\usepackage[english]{babel}
\usepackage[utf8]{inputenc}
\usepackage[T1]{fontenc}
\usepackage{bbold}
\usepackage{epsfig}
\usepackage{amsmath}
\usepackage{hyphenat}
\usepackage{hyperref}
\usepackage{placeins}
\usepackage{bm}

\begin{document}

\title{Skyrmions in square-lattice antiferromagnets}
\author{Rick Keesman}\affiliation{Instituut-Lorentz,
  Universiteit Leiden, Niels Bohrweg 2, 2333 CA Leiden, The
  Netherlands}

\author{Mark Raaijmakers} \affiliation{Institute for
  Theoretical Physics, Universiteit Utrecht, Leuvenlaan 4, 3584 CE
  Utrecht, The Netherlands}

\author{A. E. Baerends} \affiliation{Institute for
  Theoretical Physics, Universiteit Utrecht, Leuvenlaan 4, 3584 CE
  Utrecht, The Netherlands}

\author{G. T. Barkema} \affiliation{Instituut-Lorentz,
  Universiteit Leiden, Niels Bohrweg 2, 2333 CA Leiden, The
  Netherlands}\affiliation{Institute for
  Theoretical Physics, Universiteit Utrecht, Leuvenlaan 4, 3584 CE
  Utrecht, The Netherlands}

\author{R. A.  Duine} \affiliation{Institute for
  Theoretical Physics, Universiteit Utrecht, Leuvenlaan 4, 3584 CE
  Utrecht, The Netherlands}

\begin{abstract}
The ground states of square lattice two-dimensional antiferromagnets with anisotropy in an external magnetic field are determined using Monte Carlo simulations and compared to theoretical analysis. We find a new phase in between the spin-flop and spiral phase that shows strong similarity to skyrmions in ferromagnetic thin films. We show that this phase arises as a result of the competition between Zeeman and Dzyaloshinskii-Moriya interaction energies of the magnetic system. Moreover, we find that isolated (anti-)skyrmions are stabilized in finite-sized systems, even at higher temperatures. The existence of thermodynamically stable skyrmions in square-lattice antiferromagnets provides an appealing alternative over skyrmions in ferromagnets as data carriers.
\end{abstract}

\pacs{pacs}

\maketitle

\section{Introduction\label{sec1}}
Skyrmions have been the topic of intense research, in ferromagnetic materials\cite{bogdanov2,muhlbauer2,pappas1,yu2010,randeria1,han1,buhrandt1,huang2012,romming2015} as well as numerous other systems\cite{sondhi1,stoof1,leslie1,bogdanov3, ackerman2014,leonov2014}. Skyrmions in ferromagnets have promising characteristics that make them suitable for data storage and transfer: they can be driven by low critical currents\cite{muhlbauer1,sampaio1}, and they are able to move past pinning sites\cite{nagaosa1}.

Skyrmions in antiferromagnetic thin films are perhaps more suitable as data carriers than their ferromagnetic counterparts. Firstly, antiferromagnets are more prevalent in nature than ferromagnets, allowing for a wider range of material properties. Secondly, skyrmions in an antiferromagnet are less sensitive to magnetic fields. Thirdly, they move faster, and in the direction of the charge current (while skyrmions in ferromagnets experience a Magnus force with a significant component perpendicular to their trajectory), which makes it easier to control them\cite{nagaosa2,tretiakov1,ezawa1}.
For these reasons, skyrmions have been investigated in many different antiferromagnetic systems both experimentally and theoretically\cite{bogdanov1}, ranging from doped bulk materials\cite{sasagawa1}, Bose-Einstein condensates\cite{mottonen1}, various triangular lattice antiferromagnets\cite{kawamura1,pujol1}, and nanodisks\cite{ian1}.

In this manuscript we study inhomogeneous magnetization textures in square-lattice antiferromagnets (SLA's) with Dzyaloshinskii-Moriya (DM) interactions. The DM interactions that we consider arise either from bulk inversion asymmetry (symmetry class $C_{nv}$) or from structural inversion asymmetry along the thin-film normal direction. An example of the latter is an interface between a magnetic metallic system and a non-magnetic metal with strong spin-orbit coupling. For ferromagnetic systems, tunable interface-induced DM couplings have indeed been demonstrated\cite{emori1,ryu1,ryu2,franken1,je2013,hrabec2014,chen2013}. Such interfaces typically also give rise to perpendicular anisotropies, which we therefore also take into account. Finally, we also consider an external magnetic field normal to the thin film. Previous work by Bogdanov \textit{et al}.\cite{bogdanov1} considered the same system at zero temperature and in the continuum limit. These authors identified three phases: an antiferromagnetic phase, a spin-flop phase, and a phase where inhomogeneous structures persist. While examples of structures in the latter phase were given, no further phase boundaries were identified within this phase. One of our main results is that we find a distinct phase pocket that bounds a $2q$ skyrmionic phase and separates it from a spiral ($1q$) phase. Furthermore, we also confirm the existence of stable skyrmions in finite-sized systems below the Curie temperature.

The paper is organized as follows: first, we present the system under study, by defining the hamiltonian that is used in Monte Carlo (MC) simulations. After that, we discuss the various spin textures and their characteristics that arise in SLA's. We also construct the phase diagram from MC simulations, complemented by analytical results based on a continuum model. We dedicate the last two sections to the interaction energies of skyrmions, and to skyrmions in finite-sized systems, respectively, after which we conclude with a discussion and summary of our results.

\section{Model\label{sec2}}
We are interested in the equilibrium spin configurations in films of SLA materials. For this purpose, we consider a square lattice of length $L$ in the $xy$-plane with Heisenberg spins ${\bf S}_{\bf r}$ of unit length at position ${\bf r}$.
Nearest neighboring spins are coupled through an antiferromagnetic Heisenberg term $J>0$ and a Dzyaloshinskii-Moriya term $D$ and are affected by anisotropy $K$ and an external magnetic field $B$ in the ${\bf \hat{z}}$-direction. The effective hamiltonian that is used in our MC simulations is given by
\begin{align}\label{eq:hamiltoniandisc}
H =
	& J \sum_{\bf r}  {\bf S}_{\bf r}  \cdot \left(  {\bf S}_{{\bf r}+ {\bf \hat{x}}} +  {\bf S}_{{\bf r}+ {\bf \hat{y}}} \right) \nonumber \\
	+& K \sum_{\bf r}  \left( {\bf S}_{\bf r} \cdot {\bf \hat{z}}  \right)^2 - B \sum_{\bf r}  {\bf S}_{\bf r} \cdot {\bf \hat{z}} \\
	-& D \sum_{\bf r} \left(  {\bf S}_{\bf r} \times {\bf S}_{{\bf r}+ {\bf \hat{x}}} \cdot {\bf \hat{y}} - {\bf S}_{\bf r} \times {\bf S}_{{\bf r}+ {\bf \hat{y}}} \cdot {\bf \hat{x}} \right). \nonumber
\end{align}

For theoretical analysis, we consider a continuous field description of the discrete hamiltonian in Eq. (\ref{eq:hamiltoniandisc}) (see also Ref. [9]). Because of the antiferromagnetic nature of these materials, it is natural to define sublattices with magnetization ${\bf m}_1$ and ${\bf m}_2$ organized in a checkerboard configuration and put the lattice constant to unity. For antiferromagnets with large Heisenberg interaction we expect slowly varying periodic structures and the staggered magnetization ${\bf l} = ({\bf m}_1-{\bf m}_2)/2$ to be large while the total magnetization ${\bf m} = ({\bf m}_1+{\bf m}_2)/2$ is expected to be much smaller, ie, $|{\bf l}| \approx 1$ and $|{\bf m}| \ll |{\bf l}|$\cite{lifshitzbook1}. We also assume that the spatial derivatives of ${\bf m}$ can be neglected and that the contribution of the total magnetization to the anisotropic term is negligible compared to that of the staggered magnetization. This results in the following hamiltonian density
\begin{align}\label{eq:hamiltoniancont0}
\mathcal{H} =
	&\frac{J}{2} \left[ \left( \frac{\partial {\bf l}}{\partial x} \right)^2 + \left( \frac{\partial {\bf l}}{\partial y} \right)^2 + 8 {\bf m}^2\right] -B m_z + K l_z^2 \\
    &+ D \left( l_z \frac{\partial l_x}{\partial x} - l_x \frac{\partial l_z}{\partial x} + l_z \frac{\partial l_y}{\partial y} - l_y \frac{\partial l_z}{\partial y}\right). \nonumber
\end{align}
Since spins are normalized to unity, such that $|{\bf m_i}|=1$, the staggered and total magnetization must satisfy ${\bf m}^2 + {\bf l}^2=1$ and ${\bf m} \cdot {\bf l}=0$ and minimizing the hamiltonian density results in ${\bf m}=-{\bf l} \times ({\bf l} \times {\bf h})/(8J)$. Substitution leads to a hamiltonian density that is only dependent on the staggered magnetization:
\begin{align}\label{eq:hamiltoniancont}
\mathcal{H} =
	&\frac{J}{2} \left[ \left( \frac{\partial {\bf l}}{\partial x} \right)^2 + \left( \frac{\partial {\bf l}}{\partial y} \right)^2 \right] + \frac{B^2}{16 J} \left[ l_z^2 - 1 \right] + K l_z^2\\
    &+ D \left( l_z \frac{\partial l_x}{\partial x} - l_x \frac{\partial l_z}{\partial x} + l_z \frac{\partial l_y}{\partial y} - l_y \frac{\partial l_z}{\partial y}\right). \nonumber
\end{align}

We focus on systems for which the DM coupling and the coupling to the magnetic field are of the order of the Heisenberg coupling but assume that anisotropy strength is small. For small fields all higher order interactions like dipole-dipole interactions are negligible in antiferromagnets because the net magnetization is small.

\section{Phases\label{sec3}}
We find that systems described by the hamiltonian given by Eq. (\ref{eq:hamiltoniandisc}) have four distinct phases at zero temperature: the antiferromagnetic and spin-flop phase are both homogeneous, whereas the spiral and $2q$ phases are modulated, i.e., inhomogeneous phases with $1$ and $2$ dominating wave modes respectively.

\begin{center}
\begin{figure}[h] \centering
\epsfig{file=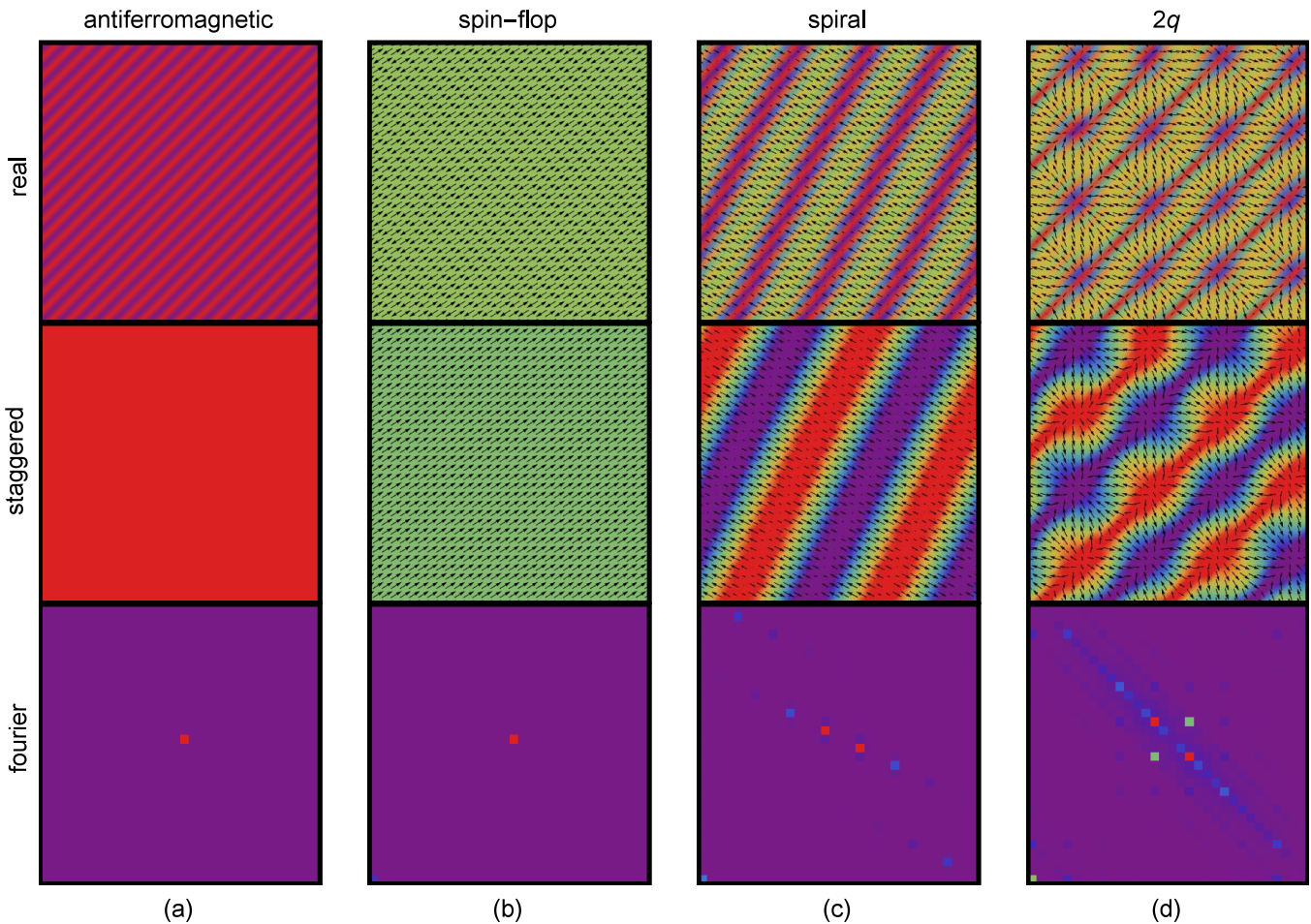,width=\linewidth,clip=}
\caption{Various types of configurations encountered in MC simulations of the model described by the Hamiltonian in Eq. (\ref{eq:hamiltoniandisc}). The antiferromagnetic (a), spin-flop (b), spiral (c), and $2q$ phase (d) are shown from left to right in typical real spin configuration (top) and staggered magnetization (middle) for an antiferromagnetic system of size $L=32$ at zero temperature. The arrows represent the local magnetization in the $xy$-plane and the background colour shows the magnetization pointing up (red) or down (purple). The norm of the Fourier modes (bottom), as defined in Eq. (\ref{eq:fourierdefinition}), of these configurations show the distinctive modes that define the phases.} \label{fig:phases1}
\end{figure}
\end{center}

There are two homogeneous phases: the antiferromagnetic phase in which the staggered magnetization points along the $z$-axis, and the spin-flop phase in which the staggered magnetization lies in the $xy$-plane. The spiral phase emerges for large enough DM coupling for which the staggered magnetization shows the same characteristics as ferromagnetic spins in a spiral state. Finally, there is a bounded region for which the $2q$ phase emerges, which has a similar texture to the spiral phase but in which the width of the spirals varies in length periodically. Configurations of these phases in real space, in terms of the staggered magnetization, and the norm of their Fourier modes are shown in Fig. \ref{fig:phases1}. The Fourier modes are defined as
\begin{align}\label{eq:fourierdefinition}
{\bf A}_{\bf q} = \frac{1}{L^2} \sum_{{\bf r}} {\bf S}_{\bf r} \exp\left[ \frac{2 \pi i}{L} {\bf q} \cdot {\bf r} \right] .
\end{align}

\onecolumngrid
\twocolumngrid

\section{Phase diagram from simulations\label{sec4}}
In one elementary move of our MC simulations, a random spin is selected and replaced by a new spin vector, drawn uniformly from a spherical cap around the original spin vector. The size of this cap is chosen such that the acceptance rate in the Metropolis algorithm is roughly $50\%$. The time step is defined such that each spin makes an elementary move once per unit of time. At each temperature typically $4500$ time steps are taken before measurements are done. During annealing, the temperature is reduced from well above the critical temperature to well below it in $200$ temperature steps. These measurements result in data obtained over a wide range of parameters and temperatures.

\begin{center}
\begin{figure}[h!] \centering
\epsfig{file=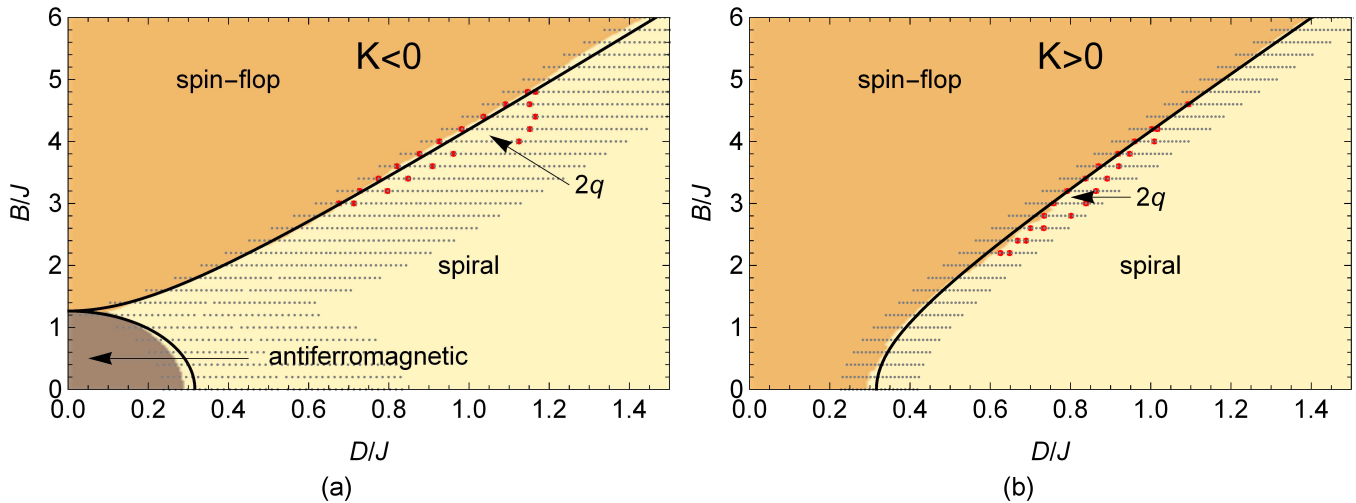,width=\linewidth,clip=}
\caption{The complete $B-D$ phase diagram for antiferromagnetic materials with (a) easy-axis anisotropy $K/J=-0.1$ and (b) easy-plane anisotropy $K/J=0.1$ at zero temperature. The grey data points display parameter values at which Monte Carlo simulations were performed. From these simulations, the phases were determined, shown as different colours. The red data points show the boundary of the $2q$ phase, as obtained from these MC simulations for fixed values of $B/J$. The analytical solutions for the phase transitions as given by Eqs. (\ref{eq:PTAFSPMINUS}-\ref{eq:PTAFSPPLUS}) are shown as solid black lines.}
    \label{fig:pd1}
\end{figure}
\end{center}

We consider the Fourier transform of the spin vectors as defined by Eq. (\ref{eq:fourierdefinition}) below. All four phases can be characterized by Fourier peaks. We define the homogeneous, spiral, and $2q$ phases as having $1$, $2$, or $4$ nonzero-mode peaks respectively. To construct the phase diagram from Monte Carlo simulations based on the discrete hamiltonian from Eq. (\ref{eq:hamiltoniandisc}) we first anneal $10$ different systems of size $L=32$ at some parameter values $J$, $D$, $B$, and $K$. From these states the one with the lowest energy is chosen, and the process is repeated for different parameter values. For all these prospective ground states the phase and the area in the phase diagram for which they have the lowest energy is determined. From this the $B$-$D$-phase diagram can be constructed for various values of anisotropy $K$. The phase diagrams are qualitatively different for systems with easy-axis ($K<0$) or easy-plane ($K>0$) anisotropy, as can be seen in Fig. \ref{fig:pd1}.

\section{Analytical phase diagram\label{sec5}}
We also construct the phase diagram by using a number of Ans\"atze for the various phases. The parameters in these Ans\"atze are obtained from minimizing the hamiltonian density from Eq. (\ref{eq:hamiltoniancont}) for these phases. This yields the transition line following Bogdanov \textit{et al}\cite{bogdanov1}.
For the antiferromagnetic phase we assume ${\bf l}=(0,0,1)$, resulting in an energy density $\mathcal{H}_{\text{AF}}=K$. The spin-flop phase is characterized by ${\bf l}=(\cos \phi, \sin \phi, 0)$ with energy density $\mathcal{H}_{\text{SF}}=-B^2 /(16J)$. For the spiral phase, ${\bf l}$ is dominated by a rotation along the direction of the wave in the $(1,1)$ direction such that ${\bf l}=\left(\sin({\bf q} \cdot {\bf r}) \cos \theta, \sin({\bf q} \cdot {\bf r}) \sin \theta, \cos({\bf q} \cdot {\bf r})\right)$. Averaging over the length of one modulation and minimizing with respect to ${\bf q}$ leads to an energy density $\mathcal{H}_{\text{SP}}=-B^2/(32J)-D^2/(2J)+K/2$. A phase transition between the antiferromagnetic and spiral, and the spin-flop and spiral phase occurs along the lines
\begin{align}\label{eq:PTAFSPMINUS}
B = 4 \sqrt{-(J K \pm D^2)} ,
\end{align}
for easy-axis anisotropy. For easy-plane anisotropy, we assume for the spiral phase that the length is also variable, i.e., ${\bf l}=l (\sin({\bf q} \cdot {\bf r}) \cos \theta, \sin({\bf q} \cdot {\bf r}) \sin \theta, \cos({\bf q} \cdot {\bf r}))$. Following the same procedure, we find that $l=\sqrt{B^2+8 D^2-8J K}/B$ minimizes the energy. The energy density for the spiral is $\mathcal{H}_{\text{SP}}=-(B^2+8 D^2-8 J K)^2/(32 B^2 J)$. In the case of easy-plane anisotropy, the phase transition between the spiral and spin-flop phase is then given by
\begin{align}\label{eq:PTAFSPPLUS}
B = 2 \sqrt{2(1+\sqrt{2})} \sqrt{D^2 - J K}.
\end{align}
We could not find such a simple Ansatz for the $2q$ phase.
\onecolumngrid
\twocolumngrid

We have constructed similar phase diagrams for various strengths of anisotropy $K/J\in{0,\pm0.02,\pm0.04,\pm0.1}$. The critical strengths $B_0$ at $D=0$ and $D_0$ at $B=0$ at which the transitions take place are obtained from Eqs. (\ref{eq:PTAFSPMINUS}-\ref{eq:PTAFSPPLUS}), yielding $B_0 \sim 4 \sqrt{|J K|}$ and $D_0 \sim \sqrt{J K}$. These become larger for increasing strengths of anisotropy. In the simulations, the size of the system limits the longest wave length of the magnetization texture. For larger anisotropy, simulations in finite sytems are therefore in better agreement with analytical calculations. An interesting point is that the $2q$ phase is always sandwiched between the spin-flop and spiral phase, at constant values of DM interaction. Its size is relatively insensitive to the strength of anisotropy. This implies that the size of modulations in the $2q$ phase in antiferromagnets is related to the pitch length $p \sim J/D$, which is a measure for the length of modulation. Therefore, $p$ only has a limited range of values, unlike the size of skyrmions in ferromagnets.

\section{Interaction energies\label{sec6}}
\begin{center}
\begin{figure}[!h] \centering
\epsfig{file=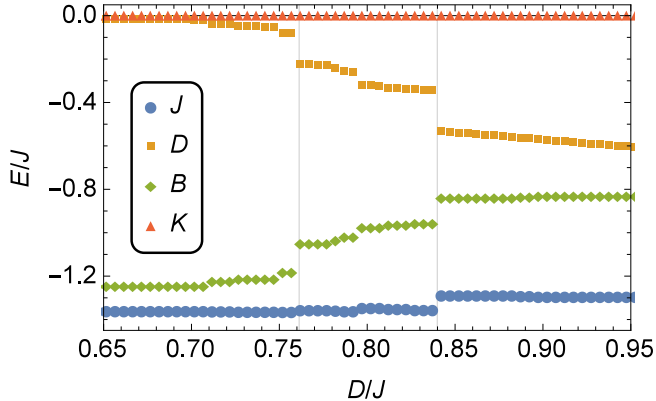,width=\linewidth,clip=}
\caption{The energy contributions $E/J$ to the ground state of the interactions with coupling parameters $J$, $D$, $B$, and $K$ are shown as a function of $D/J$ at parameter values $J=-1$, $B=3.2$, and $K=0$ for system of size $L=32$. The system undergoes two phase transitions at $D/J=0.76$ and $D/J=0.84$ between the spin-flop, $2q$, and spiral phase, respectively. These are depicted as vertical grey lines. The discrete jumps in various energy contributions suggest first-order phase transitions.} \label{fig:energy_cascade}
\end{figure}
\end{center}

To investigate further the stability of the $2q$ phase in this model, we look at the energy contributions $E$ of all interactions in the model along a line with fixed $B/J=3.2$, through the $2q$ phase in the phase diagram. With increasing DM interaction and no anisotropy, there is a transition from the spin-flop phase to the $2q$ phase at $D/J=0.76$. If the DM interaction is increased further, the spiral phase is entered at $D/J=0.84$, as can be seen in Fig. \ref{fig:energy_cascade}. At the phase transition the contributions of the external field $B$ and the DM interaction $D$ to the total energy make distinctive jumps. While the spin-flop state minimizes the energy by having a net magnetization along the external field direction, the spiral mode makes optimal use of the DM interaction. The $2q$ phase gives a compromise between the two, and so a finite area between the two emerges in which neither of them is optimal, and the $2q$ phase prevails.

\section{Skyrmions\label{sec7}}
An important question is whether the objects in the $2q$ phase as shown in Fig. \ref{fig:phases1} can be called skyrmions, as they are not fully isolated topological objects. For a ferromagnet, the (anti)skyrmion is defined as a topological object for which the winding number $w$ of the magnetization is nonzero:
\begin{align}
w =
	\frac{1}{4 \pi} \int dx dy ~ {\bf n} \cdot \left(  \partial_{x} {\bf n} \times \partial_{y} {\bf n}  \right) .
\end{align}
In case of the antiferromagnet, the winding number $w$ of the staggered magnetization can be defined instead. Although the staggered magnetization in an antiferromagnet behaves similarly as the normal magnetization in a ferromagnet there are some distinctive differences. The antiferromagnet is symmetric under the ${\bf n} \rightarrow -{\bf n}$ transformation, such that there is no difference between a skyrmion and an antiskyrmion, and neither the up or down regime in terms of the staggered magnetization is favoured over long distances. Thus there can be no lattice of isolated thermally-activated topological objects like in a ferromagnetic system, without breaking this symmetry.

\begin{center}
\begin{figure}[h!] \centering
\epsfig{file=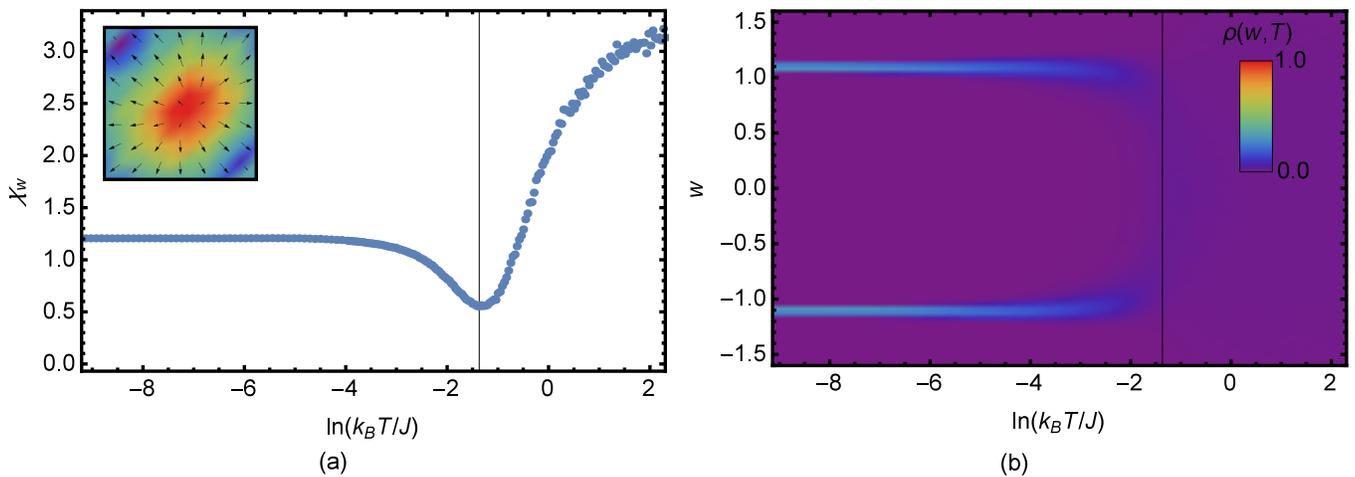,width=\linewidth,clip=}
\caption{(a) Susceptibility of the winding number of the staggered magnetization $\chi_w$ as a function of temperature $k_B T/J$, for a system of size $L=8$ with couplings $D/J=1$, $B/J=4$ and $K=0$. These parameters are chosen such that a single skyrmion emerges (see inset). The arrows represent the local N\'{e}el vector in the $xy$-plane and the background colour shows the $z$-component as positive (red) or negative (purple). A vertical line is drawn at the temperature $k_{B}T/J\approx0.25$ at which point the susceptibility is minimal. (b) Probability density $\rho(w,T)$ of the winding number $w$ of the staggered magnetization as a function of temperature. Below $k_{B}T/J\approx0.25$, indicated by a vertical line, the system chooses a configuration with either a skyrmion or an antiskyrmion.}
    \label{fig:single_skyrmion}
\end{figure}
\end{center}

In finite-size systems with open boundaries, skyrmions can, however, be stable as the boundaries break the sublattice symmetry. We find that skyrmions in finite-sized systems persist even if the temperature is increased up to the Curie temperature. To show this, we investigate a single skyrmion in a small system of size $L=8$ with open boundaries at $D/J=1$, $B/J=4$ and $K=0$, deep in the $2q$ phase (see inset of Fig. \ref{fig:single_skyrmion}(a)). The system size is chosen as the maximum size at which at most one skyrmion forms. Starting well above the critical temperature we anneal the system as discussed above. Due to sublattice symmetry, the system gets trapped in a state with either a skyrmion or an antiskyrmion in the center. From the susceptibility $\chi_w = \langle w^2 \rangle - \langle w \rangle^2$ of the winding number of the staggered magnetization, which is at a minimum at the critical temperature $\beta_c^{-1} \equiv k_{B} T_c / J$, we find $\beta_c \approx 3.9$. Results from $10^4$ annealings allow for an accurate picture of the expected number of skyrmions in this system at a certain temperature. In particular, we determine the probability density $\rho(w,T)$ for a value $w$ of the winding number at temperature $T$. Results for $\chi_w$ and $\rho(w,T)$ are shown in Fig. \ref{fig:single_skyrmion}. Within margins of error, the system contains roughly half of the time ($49.54\pm1.0\%$) a skyrmion instead of an antiskyrmion, as expected from symmetry arguments for temperatures below the critical temperature. Notice that the winding number is not exactly an integer due to edge effects. In short, this shows that (anti)skyrmions as stable isolated topological objects can exist in finite-sized systems at temperatures below the Curie temperature.
\onecolumngrid
\twocolumngrid

\section{Discussion and conclusion\label{sec8}}

In summary we have shown that certain types of antiferromagnetic thin films have four phases at zero temperature including a $2q$ phase which was not reported before. With Monte Carlo simulations and Fourier analysis, we constructed a phase diagram. The $2q$ phase has close relations to skyrmions in ferromagnetic systems, but due to symmetries a lattice of topologically isolated objects is not expected. We have shown however, that in finite-sized systems and at nonzero temperatures (anti-)skyrmions can be thermodynamically stable configurations. The existence of thermodynamically stable skyrmions in SLAs provides an appealing alternative over skyrmions in ferromagnets as data carriers.

To address finite-size effects and effects of periodic boundaries, we verified that for smaller systems of size $L=16$ the phase diagram is not significantly different. At very low $B$ and $D$ finite-size effects are stronger as long-wave-length modulated states do not fit into the small systems anymore. For parameters yielding the $2q$ phase, we verified that the conclusions presented above, which were obtained for systems of size $L=32$, still hold if the system size is increased to $L=128$. We also verified that helical boundaries with a shift up to half a period of the $2q$ phase only result in a rotation of the q-vector but otherwise do not affect the phase diagram.

Since stabilizing the $2q$ phase requires large fields $B \sim J$, the best candidates for experimental verification are antiferromagnets with low critical temperature $T_c \sim J/k_B$ so that the required fields can be more easily achieved. A possibility for experimental observation would be a monolayer of an antiferromagnetic compound that is probed by a scanning tunneling microscope, similar to the experiments with Fe\cite{romming2015} in which temperature and fields have similar energy scales.

In future work we intend to study quantum fluctuations of the ground states in the phase diagram, and how the antiferromagnetic textures interact with spin and heat current.

This work is part of the D-ITP consortium, a program of the Netherlands Organisation for Scientific Research (NWO) that is funded by the Dutch Ministry of Education, Culture and Science (OCW), and is in part funded by the Stichting voor Fundamenteel Onderzoek der Materie (FOM).

\end{document}